\documentclass{elsart}

\usepackage{epsfig}
\usepackage{amssymb}

\begin{document}

\begin{frontmatter}
\title{Prospects for Detecting a Neutrino Magnetic Moment
 with a Tritium Source and Beta-beams}
\author{G. C. McLaughlin}
\address{Department of Physics, North Carolina State University, 
Raleigh, North Carolina 27695-8202, gcmclaug@ncsu.edu}
\address{Institut de Physique Nucl\'eaire, F-91406 Orsay cedex, France}

\author{C. Volpe} 
\address{Institut de Physique Nucl\'eaire, F-91406 Orsay cedex, France,
volpe@ipno.in2p3.fr}

\begin{abstract}
We compare the prospects for detecting a neutrino magnetic moment
by the measurement of
neutrinos from a tritium source, reactors and low-energy
beta-beams.
In all cases the neutrinos or antineutrinos are detected by scattering
of electrons.
We find that a large (20 MCurie) tritium source
could improve the limit on the neutrino magnetic moment
significantly, down to the level of
a few $\times 10^{-12}$ in units of Bohr magnetons $\mu_B$
while low-energy beta-beams with sufficiently
rapid production of ions could improve the limits to the level of
a few $\times 10^{-11} \mu_B$.  The latter would require ion production at the
rate of at least $10^{15}~{\rm s}^{-1}$.
\end{abstract}


\end{frontmatter}

\maketitle

\section{Introduction}
Physicists have long speculated on the existence of a neutrino magnetic
moment. Since
recent evidence for neutrino oscillations provides indirect
evidence for a neutrino mass, this suggests that neutrinos have small magnetic
moment, although the size will depend on the nature of the neutrino mass, 
Dirac or Majorana. In the case of a Dirac mass, standard model interactions
give the neutrino a magnetic moment of 
$3 \times 10^{-19} (m_{\nu}/ {\rm eV})$ in units of Bohr
magnetons, $\mu_B$, although 
beyond the Standard Model interactions could give the neutrino
a larger magnetic moment. A positive measurement of the neutrino magnetic
moment would therefore provide valuable information for understanding
the neutrino mass mechanism. 

At present, limits on the neutrino magnetic moment come both from direct
measurements and indirect considerations. 
The most recent experiments using reactor neutrinos 
set the present upper limit of
$\mu_{\nu} < 1.0 \times 10^{-10} \mu_B$ at $90 \%$ C.L. by the MUNU
collaboration at the Bugey reactor \cite{Daraktchieva:2003dr}
and $\mu_{\nu} < 1.3 \times 10^{-10} \mu_B$ at $90 \%$ C.L. by the 
TEXONO collaboration at the Kuo-Sheng reactor \cite{texono}.
Previous measurements set similar upper limits.
From the pioneering experiment at Savannah River \cite{savannah}
and the following
experiments 
at the Kurtchatov and Rovno reactors, the limits 
$\mu_{\nu} < 2-4 \times 10^{-10} \mu_B$ \cite{Vogel:iv},
$\mu_{\nu} < 2.4 \times 10^{-10} \mu_B$ at $90 \%$ C.L. \cite{kurchatov},
$\mu_{\nu} < 1.9 \times 10^{-10} \mu_B$ at $95 \%$ C.L. \cite{rovno}  
have been derived respectively.
From solar neutrino-electron scattering at SuperKamiokande a limit of
$ < 1.5 \times 10^{-10} \mu_B$ at $90 \%$ C.L. has been obtained
\cite{Beacom:1999wx}. 
From astrophysical and cosmological considerations, namely
Big Bang nucleosynthesis, star lifetime and cooling 
and supernova explosions \cite{Raffelt:wa}, 
one infers upper limits
in the range $10^{-11}- 10^{-12} \mu_B$.
The exact values for these
limits are model dependent.
Since neutrinos mix, and different flavors
will contribute in different cases, the reactor, solar and
astrophysical limits cannot be compared directly.
However, for experiments which make use of terrestrial
neutrinos and have a short baseline relative to the energy of the
neutrinos, those originally produced as $\nu_e$ ($\bar{\nu}_e$) will
have had little opportunity to change flavor.  Whether the baseline (L)
is long or short depends on the energy of the neutrinos.  For mixing
associated with the solar neutrino solution, there
will be little transformation as long as $(L/{\rm m}) \ll 25 (E /{\rm
keV})$, and for transformation governed by the third mixing angle and
mass squared difference, the baseline is short as long as $(L/ {\rm m})
\ll 0.4 (E /{\rm keV})$.  In the latter situation
the transformation will not be large even if the baseline is long,
because the mixing angle is small and, for this reason, as yet unknown.

Since the neutrino magnetic moment interaction operates in
the mass eigenstate basis, an incoming electron neutrino
will scatter via a photon into at least three possible final states.
For Dirac neutrinos, we can write, 
 $\mu_e^2 = \sum_i |U_{ie}|^2 \mu_i^2$, where $\mu_i$ are
the magnetic moments in the mass eigenstate basis, and $|U_{ie}|^2$
are entries in the Maki-Nakagawa-Sakata (MNS) matrix.  
However, if the magnetic moment were not diagonal
the above equation should be expanded to include transition moments. 
Similarly in the case of a Majorana
neutrino, one needs to take into account a sum over all transition
moments.  In all cases we consider here however,
the measurement is of the scattering of an electron neutrino or
antineutrino into any type of neutrino by way of the magnetic
moment, so we can compare the case of reactor neutrinos, a tritium 
source and beta-beams directly without the need to decompose
a neutrino magnetic moment into its possible components in the mass
basis. 

The scenario with a tritium source that we consider is a 
200 MCurie, 20 kg source  which is used in conjunction with a large
 Time Projection Chamber (TPC) 
detector.  This has just been proposed as a way to explore 
many neutrino properties \cite{giomataris}
one of which is the neutrino magnetic moment. 
This idea in principle is very attractive 
because of the low threshold of the detector
and the large number of neutrinos from the source.

``Beta-beams'' is a novel method to produce neutrino beams which
consists in boosting radioactive nuclei, that decay through
beta-decay, to obtain an intense, collimated and pure neutrino beam 
\cite{Zucchelli:sa}.
The use of this new concept to have a
beta-beam facility for low-energy neutrinos
has been proposed by Volpe \cite{Volpe:2003fi}  
and has
the important benefit that the neutrino spectra will be
well understood. 
Interest in this method is fairly recent and 
studies of the feasibility of such beams are planned \cite{cernweb}.

In this letter we investigate the potential for setting new limits
on the neutrino magnetic moment using a tritium source or a low-energy
beta-beam. We use the specific example of a TPC detector to observe
neutrino-electron scattering. We compare the sensitivities which
could be obtained by these methods with what can be achieved with
reactor neutrinos.

\section{Calculations and Results}

Each of the sources we consider
emits a neutrino flux which peaks at different energies.  For
the tritium source and the low-energy beta-beams, the neutrino flux is
just that of a standard beta decay spectrum from a nucleus at rest.  
To calculate the spectrum, we use
 a Fermi function to approximate the final state Coulomb interaction.  
 Tritium is a low energy source, since the maximum energy of
the antineutrinos it produces is 18.6 keV.  The ions typically
considered for the beta-beams, such as $^{6}{\rm He}$, a $\beta_-$ emitter 
and $^{18}{\rm Ne}$, a $\beta_+$ emitter,
have Q-values of a few MeV, although other ions with smaller Q-values
may be considered in the future.  In table
\ref{tab:he6_spect} we
give the fluxes for these two beta-beam emitters.
Reactor  antineutrinos have a different
spectrum since they are produced by many pathways in the course of
the fissioning of the nuclei.  The understanding of
the exact nature of the reactor flux at low energies is therefore less 
certain than in the case of the beta decay of a single isotope, but
the spectrum peaks at around 1-2 MeV.   
We
use the antineutrino fluxes
as given by \cite{Vogel:iv}.

In all cases considered the neutrinos are detected through recoil electrons
from neutrino-electron scattering. The weak and electromagnetic 
cross section as a function of recoil
energy $T$ and neutrino energy $E_{\nu}$
is given by \cite{Vogel:iv}

\begin{eqnarray}
{d\sigma \over dT} & = & {G_F^2 m_e \over 2 \pi} \left[ (g_V+g_A)^2 + (g_V -g_A)^2
\left(1- {{ T \over E_{\nu}}}\right)^2 + (g_A^2 - g_V^2){m_eT \over
E_{\nu}^2}\right]   \\ \nonumber
& + & 
{\pi \alpha^2 \mu_{\nu}^2 \over{m_e^2}} {1-T/{E_{\nu}} \over T} \label{e:1}
\end{eqnarray}
where $g_V = 2 \sin^2\theta_W +1/2$, $\theta_W$ is the Weinberg angle,
  $g_A = 1/2$ (-1/2) for $\nu_e$ 
($\bar{\nu}_e$), $m_e$ is the electron mass and $G_F$ is the
 Fermi coupling constant. 
Following Eq. (1) we consider scattering of neutrino on free electrons.
The effect of atomic binding is discussed in \cite{atomic}.
These cross sections have to be averaged by the neutrino flux from the relevant
source : 
\begin{equation}\label{e:2}
\langle {d\sigma \over dT} \rangle = {{ \int { d N_{\nu} \over {d E_{\nu}}} 
{d\sigma(E_{\nu}) \over dT} dE_{\nu}} \over{\int { d N_{\nu} \over {d E_{\nu}}} 
dE_{\nu}}}
\end{equation}
where $dN_{\nu}/dE_{\nu}$ is the number of neutrinos per unit energy
emitted by the neutrino source.  Note that 
the electron recoil energy is restricted to
the values :
\begin{equation}\label{e:3}
T \lesssim {2 E_{\nu}^2 \over {2 E_{\nu} + m_e}}
\end{equation}
At lower electron recoil energies the flux-averaged weak cross section 
becomes constant
while the part proportional to the magnetic moment diverges.
Experimentally one searches for the contribution from the magnetic moment by 
looking for an excess of counts at low electron recoil energy. 
The lack of such an observation has produced the current limit on the
magnetic moment e.g. \cite{Daraktchieva:2003dr,texono,rovno}.
In tables \ref{tab:he6} and \ref{tab:ne18} we give flux-averaged electron-neutrino
cross sections (Eq.\ref{e:2}) as a function of recoil energy.

In order to consider the number of counts as a function of electron recoil 
energy
we take the specific example of a
TPC detector as in Ref.\cite{giomataris}.
 For both the beta-beams and the tritium source, 
the geometry
is the same $4 \pi$ configuration where the source is located at the
center.
The radius of the detector is 10 
meters and contains 20 tons of Xenon.
In the comparison with the reactor experiments we consider a detector
located 18 m from the source and
having 1 m$^3$ volume as used in the MUNU experiment 
\cite{Daraktchieva:2003dr}, although
we take Xenon as the detection medium.

\section{Toward New limits
on the Magnetic Moment: Tritium and Beta-Beams}

\begin{figure}[t]
\vspace*{-2cm}
\begin{minipage}{8cm}
\includegraphics[angle=0,width=6cm]{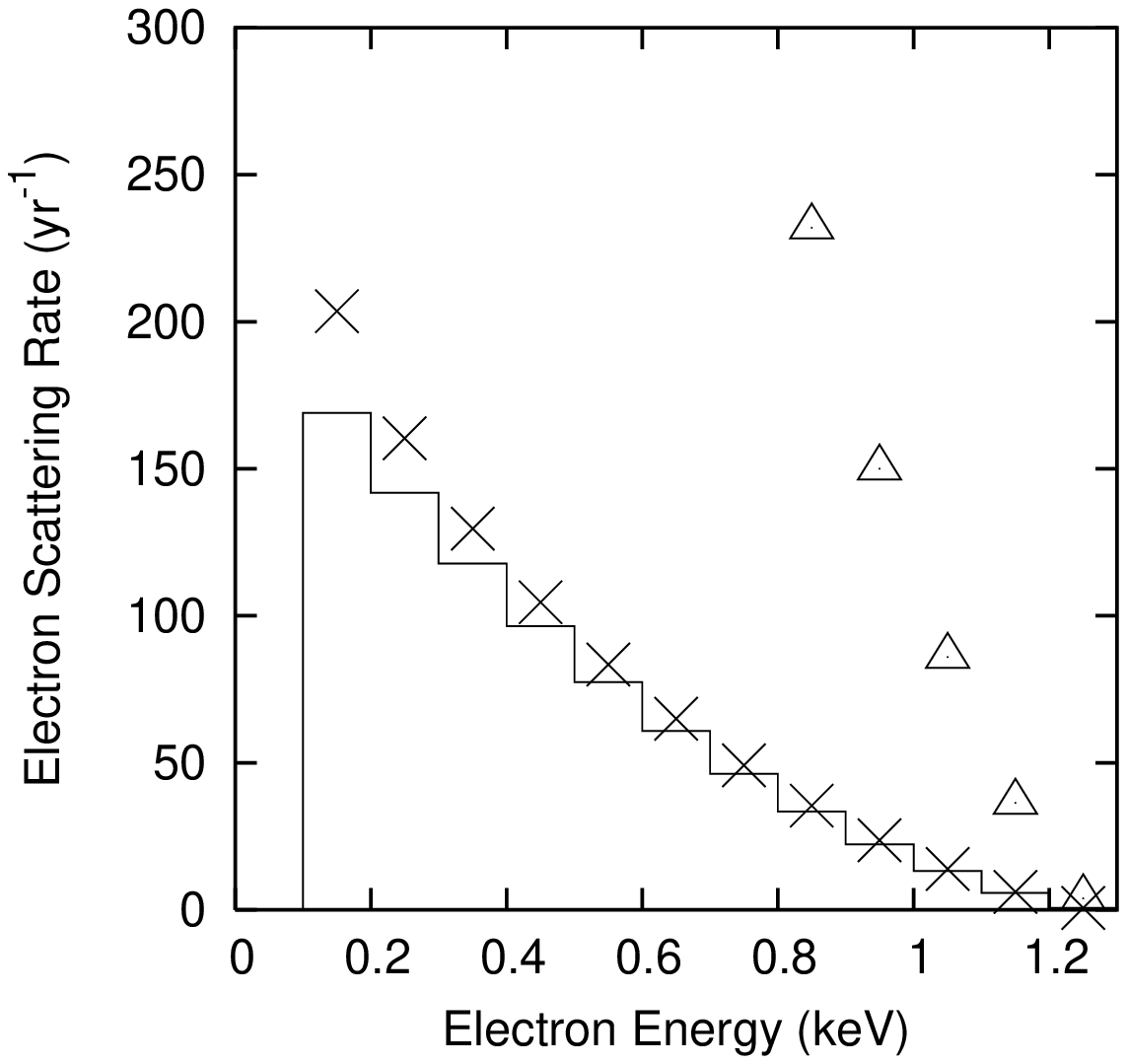}
\caption{{\sc TRITIUM SOURCE:} 
Number of neutrino-electron scattering events from a 20 MCurie Tritium
source in a 4$\pi$ TPC detector of 10m in radius.   The crosses
give the number of scatterings if the neutrino has a magnetic moment of
$\mu_\nu = 10^{-12} \mu_B$, while the triangles present the number of counts
if  $\mu_\nu = 10^{-11} \mu_B$. The histogram shows the expected number of
events for a vanishing neutrino magnetic moment.
\label{fig:trit-his}}
\end{minipage}\hspace*{0.5cm}
\begin{minipage}{8cm}
\includegraphics[angle=0,width=6cm]{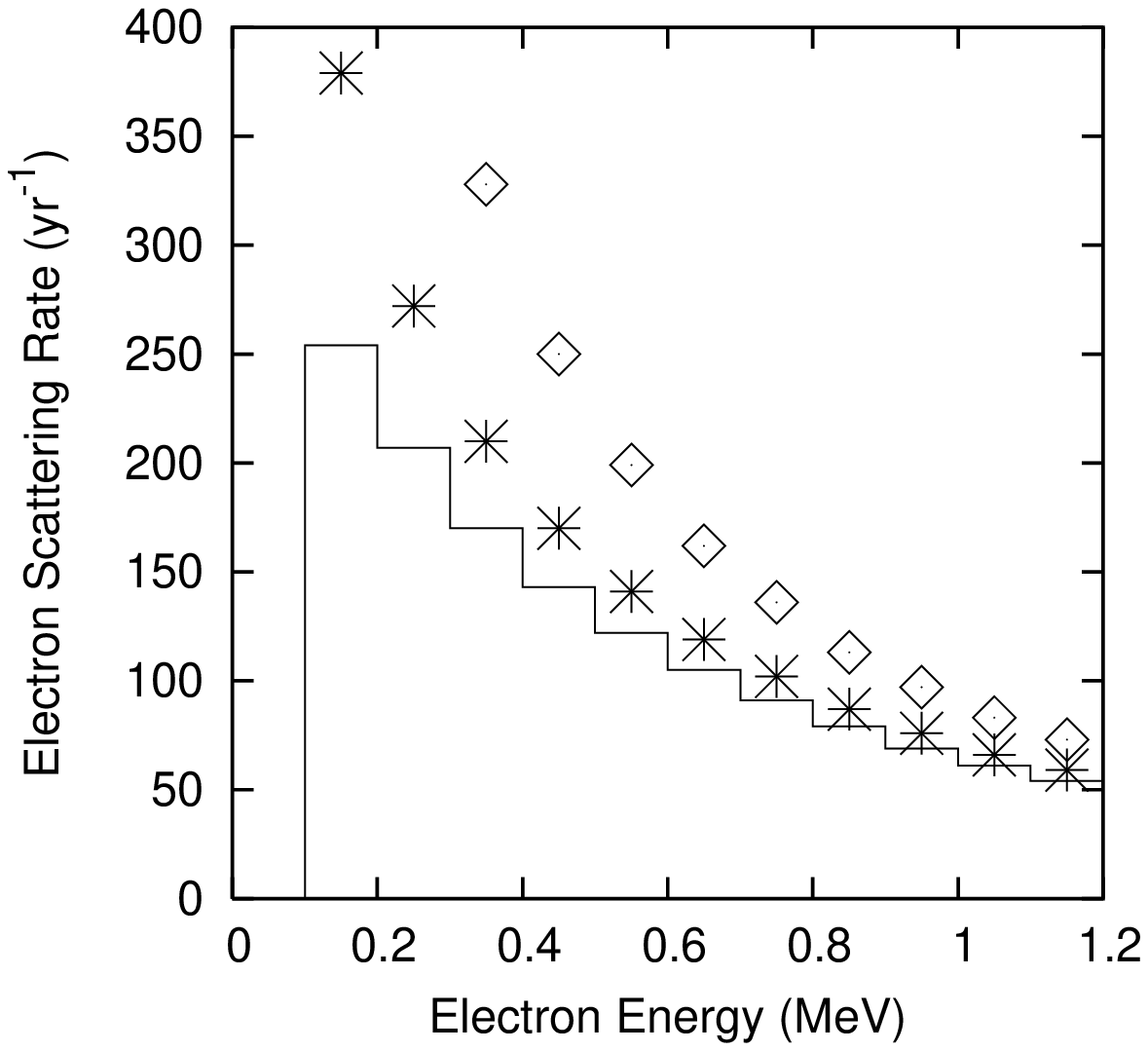}
\hspace*{0.5cm}
\caption{REACTOR:   
Number of neutrino-electron scattering events from a 2750MW 
reactor in a detector of 1m$^3$ volume at distance of 18m. The diamonds show
the number of scatterings if the neutrino has a magnetic moment of
$\mu_\nu = 10^{-10} \mu_B$, while the stars show the number of counts
if the neutrino has a magnetic moment of $\mu_\nu = 5 \times 10^{-11} \mu_B$.
The histogram shows the expected number of
events for a vanishing neutrino magnetic moment.} \label{fig:reac-his}
\end{minipage}
\end{figure}

\begin{figure}[t]
\begin{center}
\begin{minipage}{7.5cm}
\centerline{\includegraphics[angle=0,width=6cm]{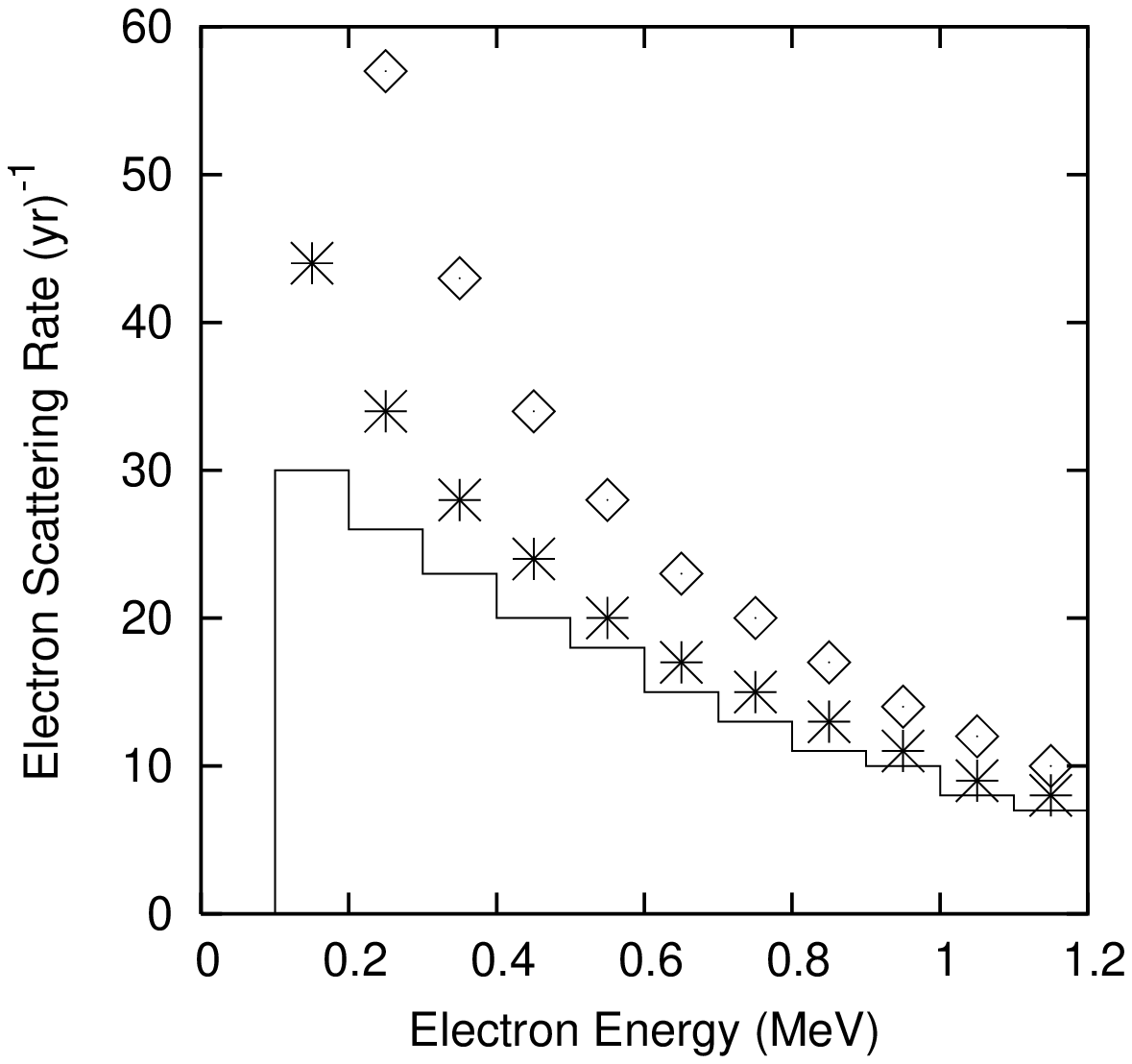}}
\end{minipage}\begin{minipage}{7.5cm}
\centerline{\includegraphics[angle=0,width=6cm]{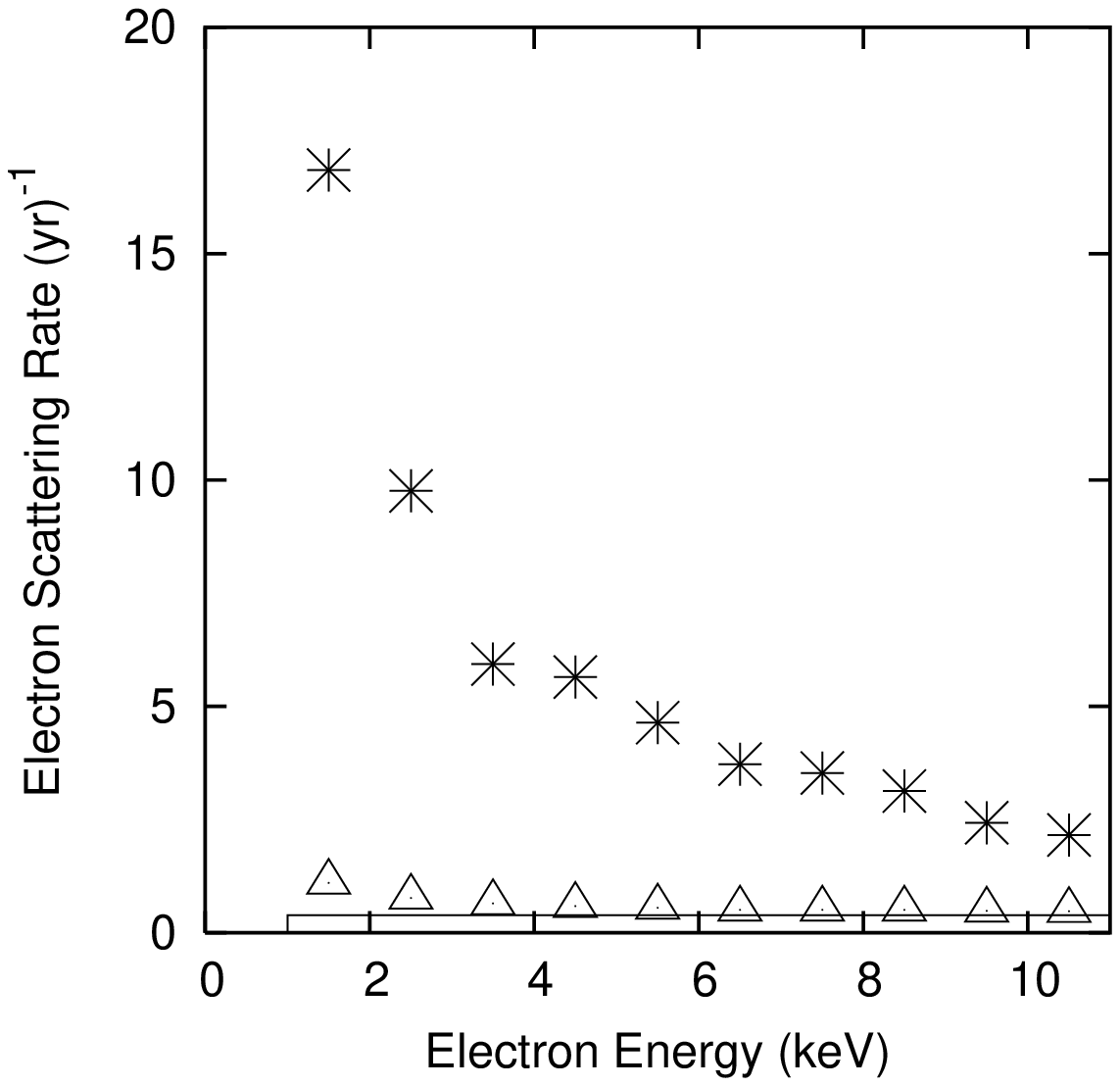}}
\end{minipage}
\end{center}
\caption{\small BETA-BEAMS: 
Number of neutrino-electron scattering events from a Helium-6
ion source produced at the rate $10^{15}$ per second, and a
$4 \pi$ detector of 10 m in radius. The diamonds show
the number of scatterings if the neutrino has a magnetic moment of
$\mu_\nu = 10^{-10} \mu_B$, the stars present the number of events
if $\mu_\nu = 5 \times 10^{-11} \mu_B$,
and the triangles give the number of events
if the neutrino has a magnetic moment of $\mu_\nu = 10^{-11} \mu_B$.
The histogram shows the expected number of
events for a vanishing neutrino magnetic moment.}\label{he6-his-kev}
\end{figure}

Here we compare the sensitivities of the recoil electron spectrum to
a non-zero magnetic moment for the three different neutrino sources.
In Figs. \ref{fig:trit-his}-\ref{he6-his-kev} 
we show the electron scattering rates as a function of
electron recoil
energy for cases with and without a neutrino magnetic moment.
The event rate is obtained by combining the flux-averaged cross sections
with both the number of electrons in the detector and the number of neutrinos
per unit time and area in the detector.
A 100 \% efficiency has been assumed in all cases. 

Figure \ref{fig:trit-his} 
shows the electron rate for the tritium source. The detector is
sensitive to an electron energy above 100 eV and the maximum electron energy
from this source is 1.26 keV. We find about 900 events/year in the absence
of a neutrino magnetic moment. For a magnetic moment of $10^{-11} \mu_B$, 
this number
is increased to ten thousand events per year.  Since the cross section
is diverging, most of these events occur at the lower energies, where the
points are so large that they are not shown on the figure. 
As can be seen from the figure the limit $\mu_{\nu}=10^{-11} \mu_B$
is clearly distinguishable while $\mu_{\nu}=10^{-12} \mu_B$, 
which would produce about one hundred events in this energy
range, seems more
challenging.  From this, one expects 
a sensitivity of a few
$\times 10^{-12} \mu_B$ to be achievable, with
 the exact value to be determined by
detailed detector simulations. 
 Fig \ref{fig:trit-his} clearly shows that the use of a 
200 MCurie tritium source,  which has a well understood spectrum coupled
with a low threshold detector, e.g.\cite{giomataris}, can be 
extremely
attractive from the point of view of setting limits on the neutrino magnetic
moment. 

Scattering rates for neutrinos produced by reactors are shown in
Figure \ref{fig:reac-his}.
For the purposes of this figure,
 we assume a 2750 MW reactor, like the Bugey reactor,
and a fuel composition of $55.6 \%$ $^{235} {\rm U}$, $32.6 \%$
$^{239} {\rm Pu}$, $7.1 \%$ $^{238} {\rm U}$ and $4.7 \%$
$^{241} {\rm Pu}$.
For these values the number of neutrinos produced is about
$4 \times 10^{20} \nu/ {\rm s}$ and  as stated before, we assume
  that
the detector is located
about 18 meters away  from the source.
Since the reactor neutrino flux peaks at 1-2 MeV there are significant
electron recoil events in the tenths of MeV range.
Between 0.1 MeV and 1.2 MeV, there would be about 400 events in three
 months with this
detector and this number would increase to 500  with a neutrino
magnetic moment of $5 \times 10^{-11} \mu_B$.  Therefore,
if the electron recoil is measured in this range, it is
in principle sensitive to a magnetic moment of $5 \times 10^{-11} \mu_B$
or even  smaller. In order to reach this it is crucial to understand
the low energy part of the neutrino spectrum \cite{Li}.
In fact, there are several sources of uncertainty, like for example
$\beta$-decays, which are not well known,
either of the fission products or induced by
neutrons absorbed by the materials.
A detailed discussion of the influence of such uncertainty
on the $\mu_{\nu}$ sensitivity, for different energy ranges of the 
electron recoil energy, is presented in \cite{Li}.
The knowledge of the low energy spectra is a limiting
factor in the MUNU experiment \cite{Daraktchieva:2003dr} which quotes a limit
based on data taken above 900 keV.

Let us now consider the case of a beta-beam source.
Similarly to the case of a static tritium source, an advantage of the
beta-beams is that the neutrino fluxes can be very accurately calculated.
Figure \ref{he6-his-kev} shows 
the electron-neutrino scattering events in the
range of  0.1 MeV to 1 MeV and 1 keV to 10 keV respectively.
(In Figure 3b we have rounded to the nearest integer number of counts).
The shape of the flux-averaged cross
sections is very similar to the reactor case as reflected in the event
rates shown in the figures. As can be seen, by measuring
electron recoils in the keV range with a beta-beam source one
could, with a sufficiently strong source, have a very clear signature for     
a neutrino magnetic moment of $5 \times 10^{-11} \mu_B$.
These figures are for Helium-6 ions, however, 
similar results can be obtained using neutrinos from $^{18} {\rm Ne}$.  
The results shown are obtained for an intensity of $10^{15}~\nu/s$ (i.e.
$10^{15}~$ ions/s).  If there is no magnetic moment, this intensity
will produce about 170 
events in the 0.1 MeV to 1 MeV range
per year and  3 events in the 1 keV to 10 keV range per year.  
These numbers increase to 210 and 55 respectively in the case of a 
magnetic moment of  $5 \times 10^{-11} \mu_B$.

Although the first feasibility study shows that an intensity of
a few $\times 10^{13}~ {\rm ions/s}$ can be reached for $^6{\rm He}$ 
\cite{feas}, 
higher intensities such as $10^{15}~
{\rm ions/s}$ may be obtained for specific ions and/or designs
 \cite{pcmats}.
An intensity at least that high is needed to 
improve the present limit on the neutrino magnetic moment.

\section{Conclusions}
We have discussed different routes for obtaining improved limits
on the neutrino magnetic moment.  The best scenarios for
improving the limits, or indeed for detecting a neutrino magnetic
moment, will involve three main aspects.  One needs to understand
well the spectrum of the neutrino source, to be able to measure
electron recoil at as low energy as possible, and to have an
intense neutrino flux.

The main advantage of a large
tritium source when combined with a low threshold detector such as
a large TPC detector, is that the neutrino flux is well known and many
counts can be obtained at low recoil energy.  Such a configuration
is very attractive since one can obtain sensitivities to the neutrino 
magnetic moment of a few $\times 10^{-12} \mu_B$. 

Reactor
neutrinos have provided the best terrestrial limits on the neutrino
magnetic moment to date $\mu_\nu \lesssim 10^{-10} \mu_B$.  
Reactor experiments  have the advantage that there is an intense source that
can be utilized.  In order to improve the current limits using 
reactor neutrinos,   one needs to be able to measure and 
understand well the
electron recoil spectrum below an MeV, for which it is necessary
to understand the low energy component of the neutrino flux.

Beta-beams provide a new and as yet unexplored possibility for
measuring a neutrino magnetic moment.  As in the case of the
tritium source, the knowledge of the spectrum of neutrinos
is very good. The limiting feature of this scenario is
the intensity of the neutrino fluxes.  One needs at least 
$10^{15}~ {\rm ions/s}$
in order to improve the current limit and reach a~few~$\times 10^{-11} \mu_B$. 
For higher fluxes, one
obtains better limits. For a low energy application such as this, 
it would be extremely helpful to evaluate ways 
to attain high production rates for beta-beam ions.  

\vspace{.3cm}
We thank J. Bouchez, I. Giomataris, M. Lindroos, A. Villari, H. Weick
 for useful discussion.
G.C.M. thanks IN2P3 for their support while this work was being completed
 and also acknowledges support from DOE Grant DE-FG02-02ER41216.

\newpage

\begin{table}[h]
\begin{minipage}{8cm}
\begin{tabular}{|c|c|} \hline
~~ E$_{\bar{\nu}}$ (MeV)~~ & ~~${dN_\nu \over dE_\nu}$~~ \\
 \hline
  0.35 & 0.04 \\
  0.70 & 0.14 \\
  1.1  & 0.27 \\
  1.4  & 0.36 \\
  1.75 &  0.40 \\
  2.1 &  0.41 \\
  2.5 &  0.37 \\
  2.8 &  0.28 \\
  3.2 &  0.14 \\
  3.5 &  0.02 \\
\hline
\end{tabular}

\end{minipage}\hspace*{0.5cm}
\begin{minipage}{8cm}
\begin{tabular}{|c|c|} \hline
~~ E$_{\nu}$ (MeV)~~ & ~~$dN_\nu \over dE_\nu$~~ \\
 \hline
  0.34 &  0.02 \\
  0.68 & 0.06 \\
  1.0   & 0.11 \\
  1.4  & 0.16 \\
  1.7  & 0.18 \\
  2.1  & 0.19 \\
  2.4  & 0.17 \\
  2.7  & 0.13 \\
  3.1  & 0.07 \\
  3.4  & 0.01 \\
\hline
\end{tabular}\end{minipage}
\caption{The left table shows the spectrum of antineutrinos 
(${\rm MeV}^{-1} \; {\rm s}^{-1}$) from $^{6}$He and the 
right table shows the spectrum of neutrinos (${\rm MeV}^{-1} \; 
{\rm s}^{-1}$) from $^{18}$Ne from the decay to the ground state 
of $^{18}$F (92\% branch). \label{tab:he6_spect}}

\end{table}
 
\begin{table}[h]
\begin{tabular}{|c|c|c|} \hline
~~ T recoil (MeV)~~ & ~~{\sc weak}~~ & ~~{\sc magnetic}$\times
(\mu_{\nu}/10^{-10})^2$~~ \\
 \hline
$3.5 \times 10^{-4}$ & 10.1  & 7090. \\
$7.0 \times 10^{-4}$ & 10.1  & 3540. \\
$1.0 \times 10^{-3}$ & 10.1 & 2360. \\
$5.3 \times 10^{-3}$ & 10.1  & 471.\\
$1.0 \times  10^{-2}$ & 10.0 & 243. \\
$5.0 \times 10^{-2}$ & 9.5  & 48. \\
$1.0 \times 10^{-1}$ & 8.9  & 23. \\
$5.0 \times 10^{-1}$ & 5.2  & 3.4 \\
1.0 & 2.5  & 1.0 \\
3.1 &  6.0 $\times 10^{-3}$ & 7.2 $\times 10^{-4}$ \\ \hline
\end{tabular}
\caption{Flux-averaged cross sections,
$\langle d\sigma / dT \rangle$ in units of ($10^{-45} \; {\rm cm}^2
{\rm MeV}^{-1}$)  
as a function of electron recoil
for electron scattering of $^{6}$He antineutrinos.
The column labeled {\sc weak} 
was calculated with the term in Eq.(\ref{e:1})
which is  proportional
to $G_F^2$, while the column labeled {\sc magnetic} 
was calculated with the
term in Eq.(\ref{e:1}) which is proportional to $\mu_\nu^2$.
\label{tab:he6}}
\end{table}

\begin{table}[h]
\begin{tabular}{|c|c|c|} \hline
~~ T recoil (MeV)~~ & ~~{\sc weak}~~ & ~~{\sc magnetic}$\times
(\mu_{\nu}/10^{-10})^2$~~ \\
 \hline
$3.4 \times 10^{-4}$ & 10.1 & 7260.\\
$6.8 \times 10^{-4}$ & 10.1 & 3630. \\
$1.0 \times 10^{-3}$ & 10.1 & 2420. \\
$5.1 \times 10^{-3}$ & 10.1 & 483. \\
$1.0 \times  10^{-2}$ & 10.1 & 240. \\
$5.0 \times 10^{-2}$ & 10.0 & 48. \\
$1.0 \times  10^{-1}$ & 9.9 & 23. \\
$5.0 \times 10^{-1}$ & 8.9 & 3.4 \\
1. & 7.3  & 1.0 \\
3.1 &  2.5 $ \times 10^{-2}$  & 1.8$\times 10^{-4}$ \\ \hline
\end{tabular}
\caption{ As in Table \ref{tab:he6}, 
flux-averaged cross sections ($10^{-45} \; {\rm cm}^2{\rm MeV}^{-1}$)
as a function of electron recoil
for electron scattering of $^{18}$Ne neutrinos are shown.\label{tab:ne18}}
\end{table}

\end{document}